\documentclass{appolb}
\usepackage{graphicx}
\usepackage{lineno}

\begin{document}
\title{Study of path-length dependent energy loss of
jets in p--Pb and Pb--Pb collisions with ALICE
\thanks{Presented at the 29th International Conference on Ultrarelativistic Nucleus-Nucleus Collisions}%
}
\author{C. Beattie for the ALICE Collaboration
\address{Yale University}
}

\maketitle
\begin{abstract}
Jet quenching, a standard signature of quark--gluon plasma (QGP) formation in which jets lose energy by traversing the medium, comprises a well-studied set of observables in heavy-ion collisions. Significant questions remain, however, concerning the mechanisms driving this phenomenon. Theoretical work that attempts to address these open questions offers the path-length dependence of jet energy loss as one way to better understand the underlying mechanisms of jet quenching. It has proven challenging, however, to derive explicit values for the path-length dependence from experimental data. These proceedings discuss recent results from ALICE that attempt to contribute to our understanding of this phenomenon, including results of event-shape engineered jet spectra, the jet-particle $v_{2}$, and correlation studies between hard triggers and hadrons.
\end{abstract}

\section{Introduction}
It is well-established that jets lose energy as they traverse the quark--gluon plasma (QGP). There is much, however, that is still unknown about the processes by which this occurs. For example, the path-length dependence of jet energy loss has not been experimentally constrained. Such constraints would enable, however, a better understanding of the relative contributions of the energy loss mechanisms that high-$p_{\rm T}$ partons may undergo. This is because collisional energy loss is expected to scale linearly with the path-length traversed, whereas radiative energy loss is expected to scale quadratically (assuming a static medium in the weakly coupled limit) \cite{djordjevic}. This theoretical relationship has motivated many experimental efforts to quantify the aforementioned dependence, which has thus far proven difficult to do for several reasons, including the fluctuating nature of parton energy loss. Given these challenges, novel methods that correlate the soft and hard activity of an event hold promise for elucidating this dependence in a way that has been to this point inaccessible.\par
In practice, this is typically done by considering jets and jet-like objects and their event plane dependent behaviour, where the event plane is taken to be the experimental estimate of the reaction plane \cite{eventplane}. Consider, for example, a semicentral collision with an almond-shaped overlap. In such a collision, it is expected that the in-plane axis will be shorter than the out-of-plane axis. It is accordingly expected that the energy loss of a hard parton will be greater along the out-of-plane axis if path-length dependence is a leading contributor to energy loss. Studying observable modification as a function of this angle therefore offers an approach to understanding the driving mechanisms of jet energy loss. These proceedings will discuss the current efforts at ALICE to measure such observables and contribute to our overall understanding of jet quenching. 

\vspace{-2mm}

\section{Results}

The first type of measurement that is considered here is an Event-Shape Engineering (ESE) study. ESE allows for the selection of events based on their ellipticity within a centrality class \cite{ese}. This is done using the magnitude of the second-order harmonic reduced flow vector, $q_{\rm 2}$, as measured in the V0C detector (covering the pseudorapidity range $ -3.7 < \eta < -1.7 $), where $q_{\rm 2}$ is defined as 
\begin{equation}
    q_{2} = |(\sum^{M}_{i = 1}  \cos(2\varphi_{\rm i}), \sum^{M}_{i = 1} \sin(2\varphi_{\rm i}))|/\sqrt{M},
\end{equation}
where $\varphi_{i}$ is the azimuthal angle of the $i^{\rm th}$ particle and $M$ is the measured multiplicity. After calculating $q_{\rm 2}$, events can be sorted into large and small ellipticity classes, where small $q_{\rm 2}$ events are highly isotropic and large $q_{\rm 2}$ events are highly elliptical in shape. When comparing the charged jet spectra from these classes, it can be seen that the ratio of the spectra is compatible with unity, as shown in Fig. 1 (left). This result suggests that a fully corrected jet measurement is relatively insensitive to the radial flow of the underlying event. By further incorporating the jets' dependence on the event plane angle, one can study the question of path-length dependence described above. It is expected that very elliptical events will have more extreme differences between in- and out-of-plane path-lengths than very round events, and therefore a more extreme difference in energy loss between in- and out-of-plane jets in the case where path-length dependence is a significant contributor to energy loss \cite{trajectum}. It is shown in Fig. 1 (right) that, for the low- to mid-transverse momentum ($p_{\rm T}$) range, out-of-plane jets are more suppressed than in-plane jets in high $q_{\rm 2}$ events than low $q_{\rm 2}$ events, corresponding with the expectation of increased suppression. Note that while the $q_{2}$ values used in this measurement were obtained from the V0C, events in which the V0A activity would indicate an opposite classification were rejected to minimize the contribution from auto-correlations. Additionally, these results are corrected for the event plane resolution using the three sub-event method \cite{eventplane}.

\begin{figure}[htb]
\centerline{%
\includegraphics[width=0.5\textwidth]{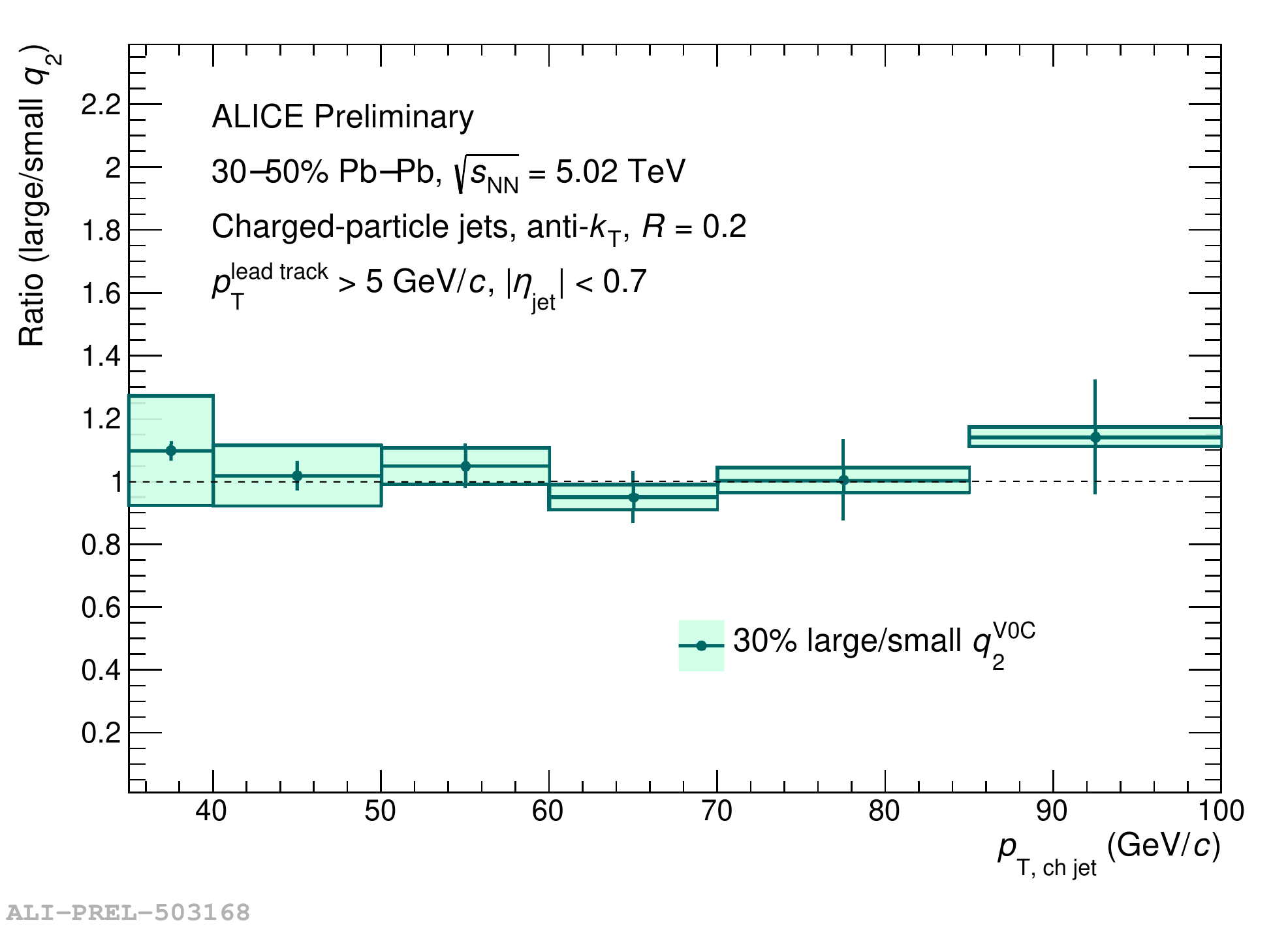}
\includegraphics[width=0.5\textwidth]{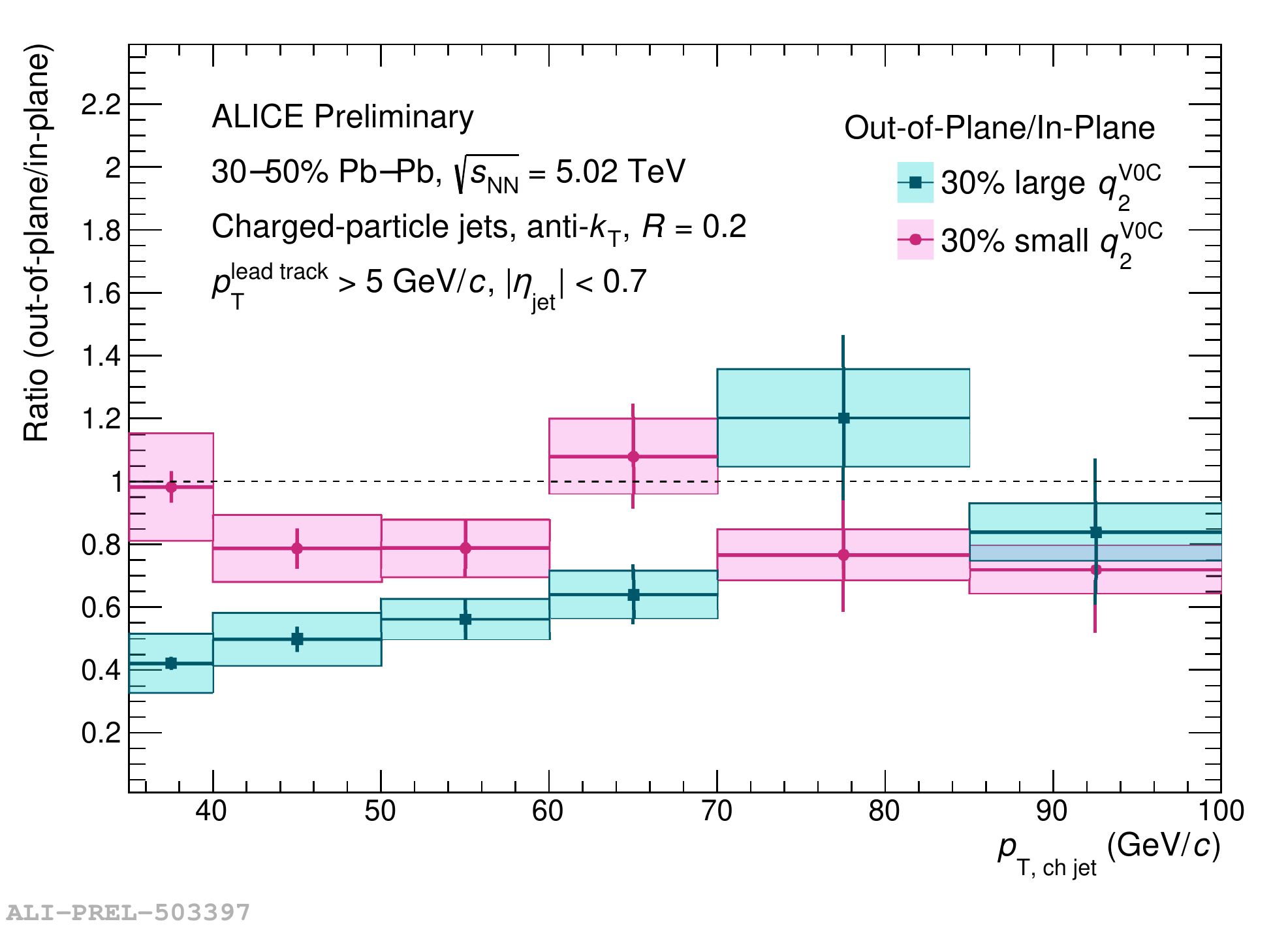}}
\caption{On the left is shown the ratio of jet spectra for large and small $q_{2}$ events. On the right is the ratio of out-of-plane to in-plane jet spectra for large (blue) and small (fuchsia) $q_{\rm 2}$ classes in Pb--Pb collisions at $\sqrt{s_{\rm NN}} = 5.02$ TeV. }
\label{Fig:F2H}
\end{figure}

Another novel measurement that explores this topic is the jet-particle $v_{\rm 2}$. The jet-particle $v_{\rm 2}$ measurement is analogous to the jet $v_{\rm 2}$, but considers the particles associated with the hard scatter as opposed to that of the reconstructed jet. This is done by extracting jet signals using double Gaussian distributions fitted to two-particle correlations with a hard trigger in the Time Projection Chamber (TPC), and then subtracting a sum of harmonics as the background. Pairs of these TPC particles are then correlated with particles from a forward detector to form a three-particle correlation and measure the $v_{\rm 2}$ \cite{jpv2}. Only same-sign particles are considered within the TPC to suppress contributions from resonance decays. Fig. 2 shows the jet-particle $v_2$ measured in p--Pb and Pb--Pb collisions at $\sqrt{s_{\rm NN}} = 5.02$ TeV. Non-zero, positive jet-particle $v_2$ values are observed throughout the entire $p_{\rm T}$ range in Pb--Pb collisions. This $v_2$ agrees with the reconstructed jet $v_2$ at high $p_{\rm T}$, while its magnitude is less than that of inclusive particles at low $p_{\rm T}$. This difference suggests that a different underlying mechanism is driving the anisotropy of jet particles than that of bulk particles at low $p_{\rm T}$. Interestingly, a non-zero $v_{\rm 2}$ can also be seen for p--Pb collisions. While this signal is notably smaller in p--Pb than in Pb--Pb, it raises the question of whether jet quenching is responsible for the anisotropy in both cases, or if a different mechanism is at play. 

\begin{figure}[htb]
\centerline{%
\includegraphics[width=0.7\textwidth]{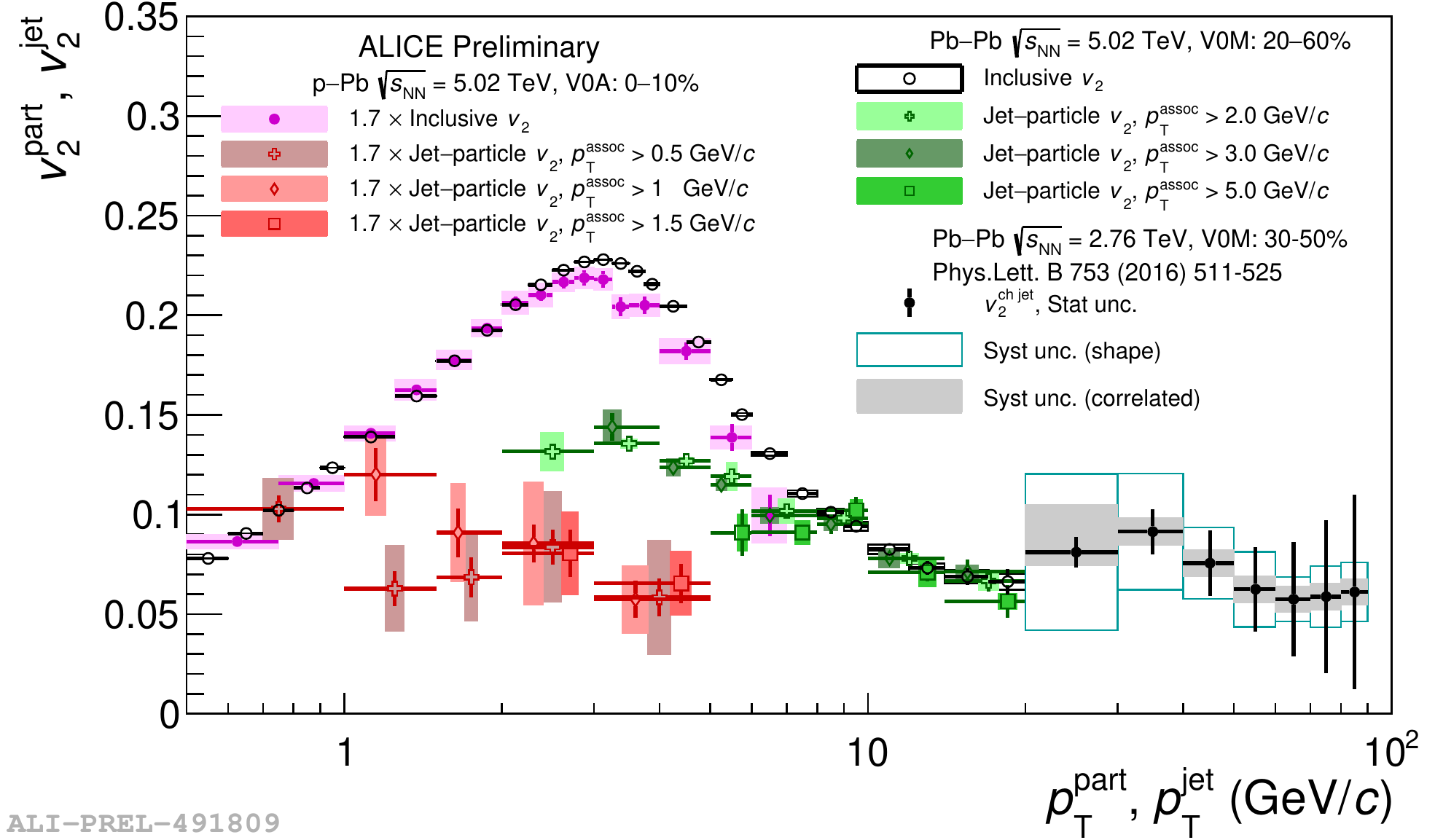}}
\caption{Jet-particle $v_2$ compared with inclusive-particle $v_2$ and jet $v_2$~\cite{jetv2} in p--Pb and Pb--Pb collisions at $\sqrt{s_{\rm NN}} = 5.02$ TeV. }
\label{Fig:F2H}
\end{figure}

A third class of measurements that attempts to address the path-length question consists of event plane dependent correlation studies. These measurements consider a high-$p_{\rm T}$ trigger and measure the yields of particles associated to this trigger using the Reaction Plane Fit method \cite{rpf}. How these associated yields change with the event plane dependence of the trigger is proposed as a useful way to understand path-length dependent modification of hard probes, and can be studied for a variety of trigger types. \par

The first such correlation measurement presented here is the $\pi^{0}$-hadron correlation. This measurement identifies trigger pions between 11 and 14 GeV/\textit{c} using the ALICE Electromagnetic Calorimeter \cite{EMCal}. Associated particle yields are considered on both the near- and away-sides. Measuring the ratio between yields associated with pions out-of- and in-plane, as shown in Fig. 3, it can be seen that there is no significant deviation from unity. Notably, JEWEL (without recoils) does not predict any such modification \cite{JEWEL}. This result is not immediately intuitive, and future work, such as incorporating JEWEL with recoils, is needed to clarify the interpretation. A second correlation-type measurement is the jet-hadron correlation. In this case, a jet with $20<p_{\rm T}<40$ GeV/$c$ is used as the trigger. Associated particle yields on both the near- and away-sides are again measured, as in the $\pi^{0}$-hadron correlation measurement. In contrast to the $\pi^{0}$-hadron results, a significant deviation from unity is observed at low-$p_{\rm T}$ on the away-side, as seen in Fig. 4. Moreover, this is inconsistent with JEWEL calculations (again, without recoils). \par

\begin{figure}[htb]
\centerline{%
\includegraphics[width=0.5\textwidth]{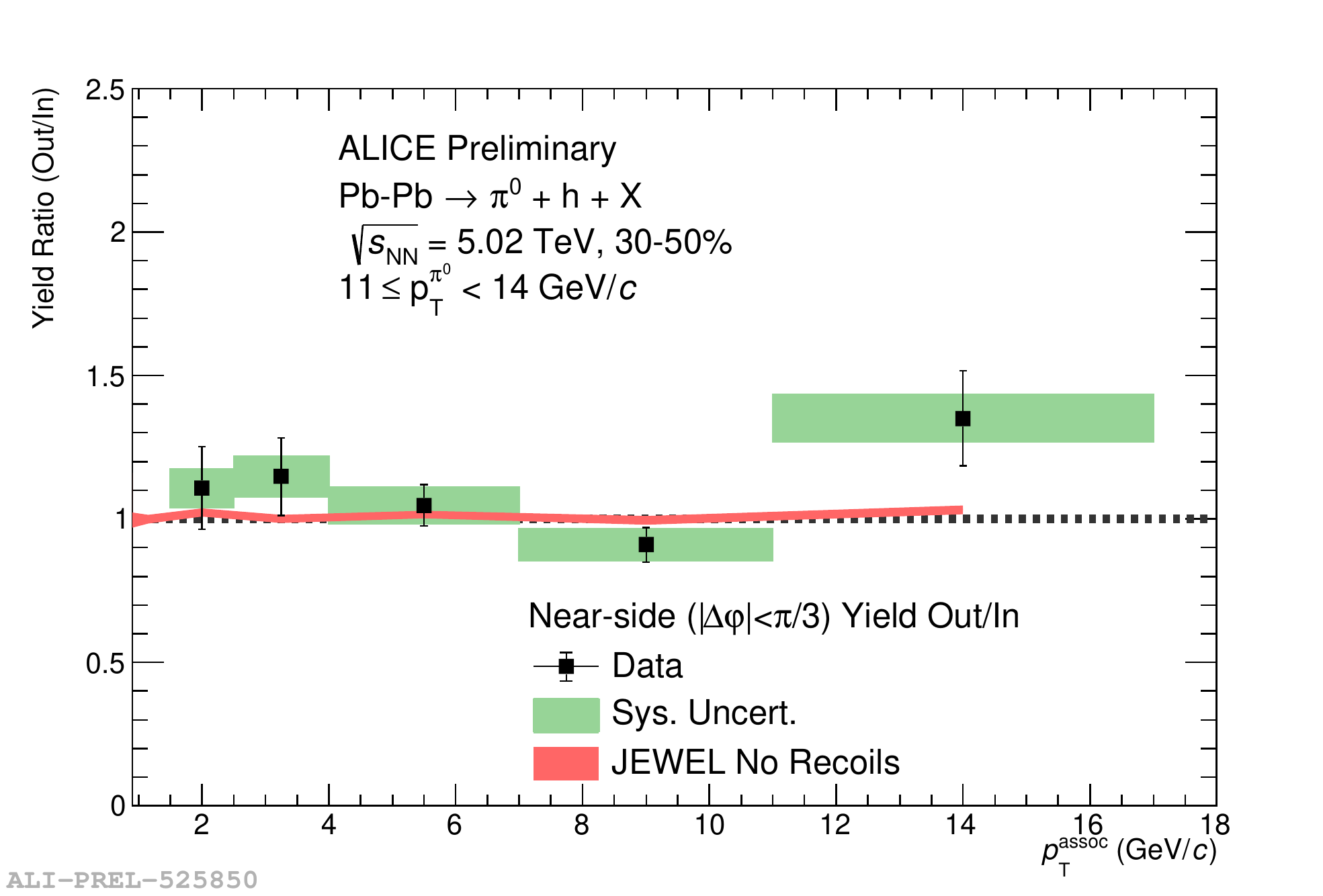}
\includegraphics[width=0.5\textwidth]{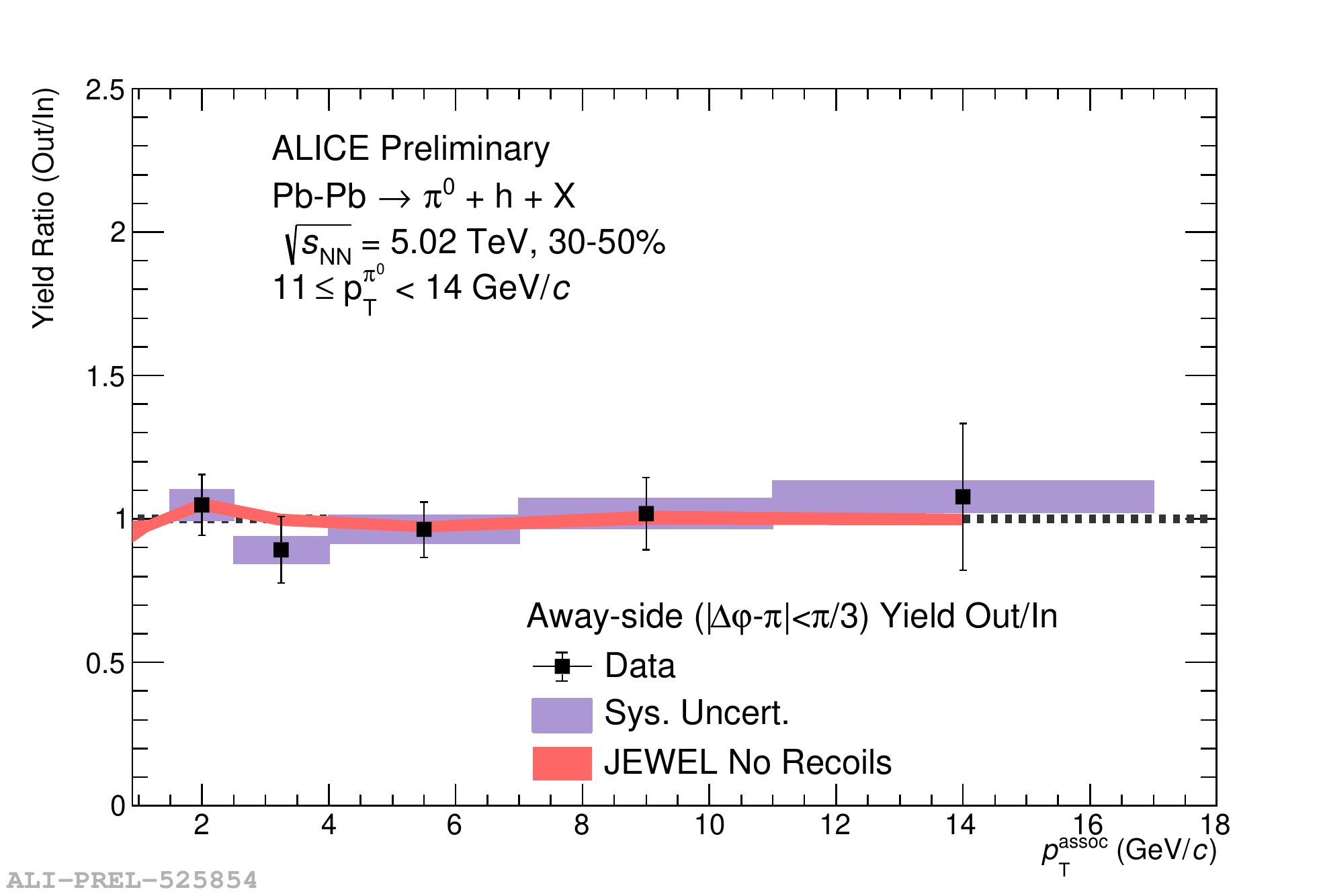}}
\caption{Near-side (left) and away-side (right) $\pi^0$-hadron correlation yield ratios for out-of-plane to in-plane associated hadrons in Pb--Pb collisions at $\sqrt{s_{\rm NN}} = 5.02$ TeV compared with JEWEL calculations.}
\label{Fig:F2H}
\end{figure}

These inconsistencies raise significant questions as to how such measurements should be interpreted. One possible explanation for the discrepancy comes from the fluctuating nature of jet energy loss. An initiating parton with a given $p_{\rm T}$ does not lose a predetermined amount of energy as it traverses the QGP, but rather loses energy according to a probabilistic distribution that depends on how the parton fragments, fluctuates in the medium, etc. This already introduces complexity into the interpretation of inclusive jet measurements. Consider, then, that in the case of correlation studies with two independently fluctuating objects, any dependence on path-length can become a sub-leading effect. It has been shown, for example, that the dijet asymmetry can be reproduced in JEWEL purely by implementing fluctuations in jet energy loss and without accounting for any path-length dependent contributions \cite{zapp}. In these correlation measurements, then, it is important to understand and account for a wide range of effects that can introduce seeming discrepancies depending on the chosen trigger.

\begin{figure}[htb]
\centerline{%
\includegraphics[width=0.5\textwidth]{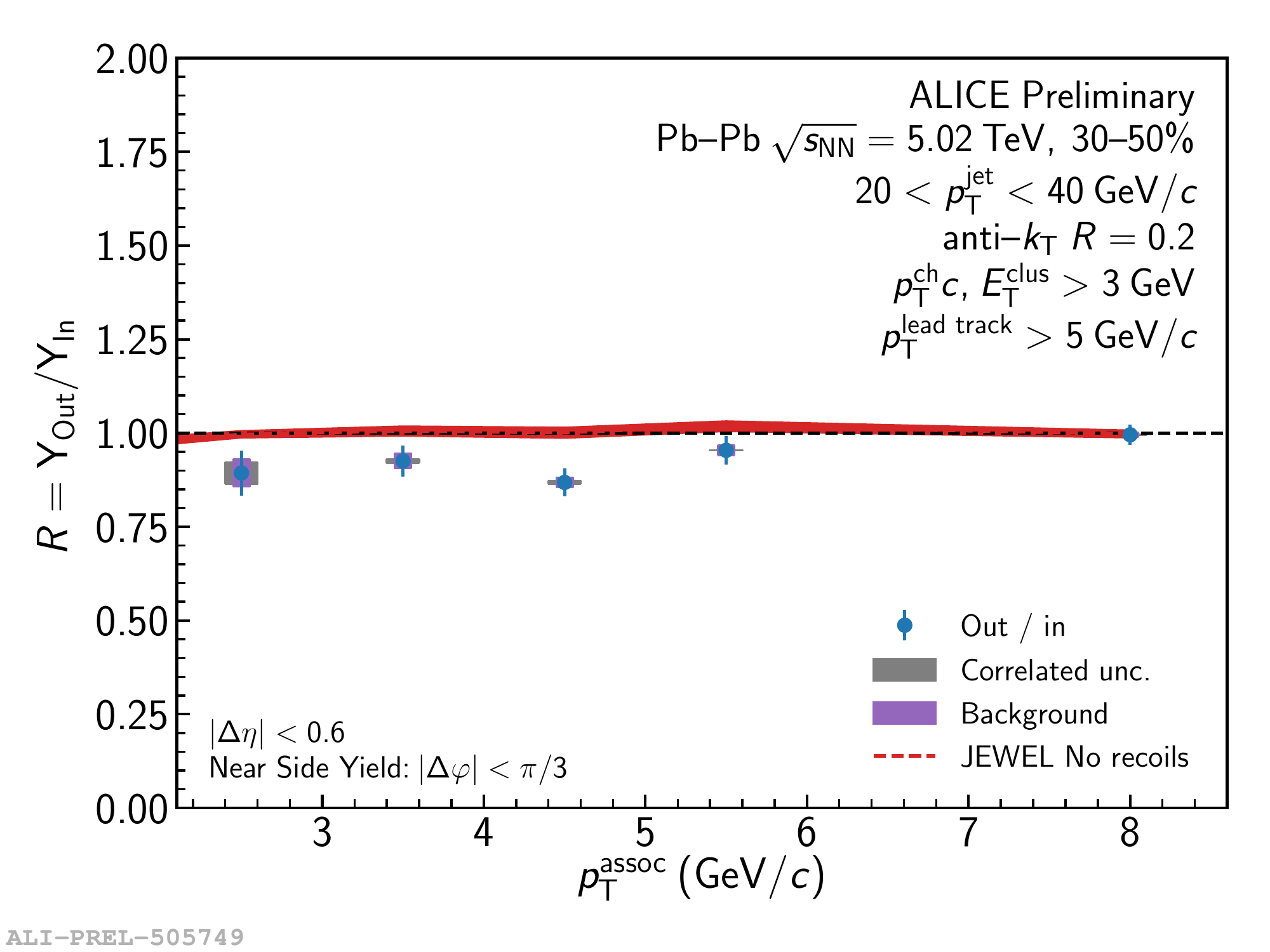}
\includegraphics[width=0.5\textwidth]{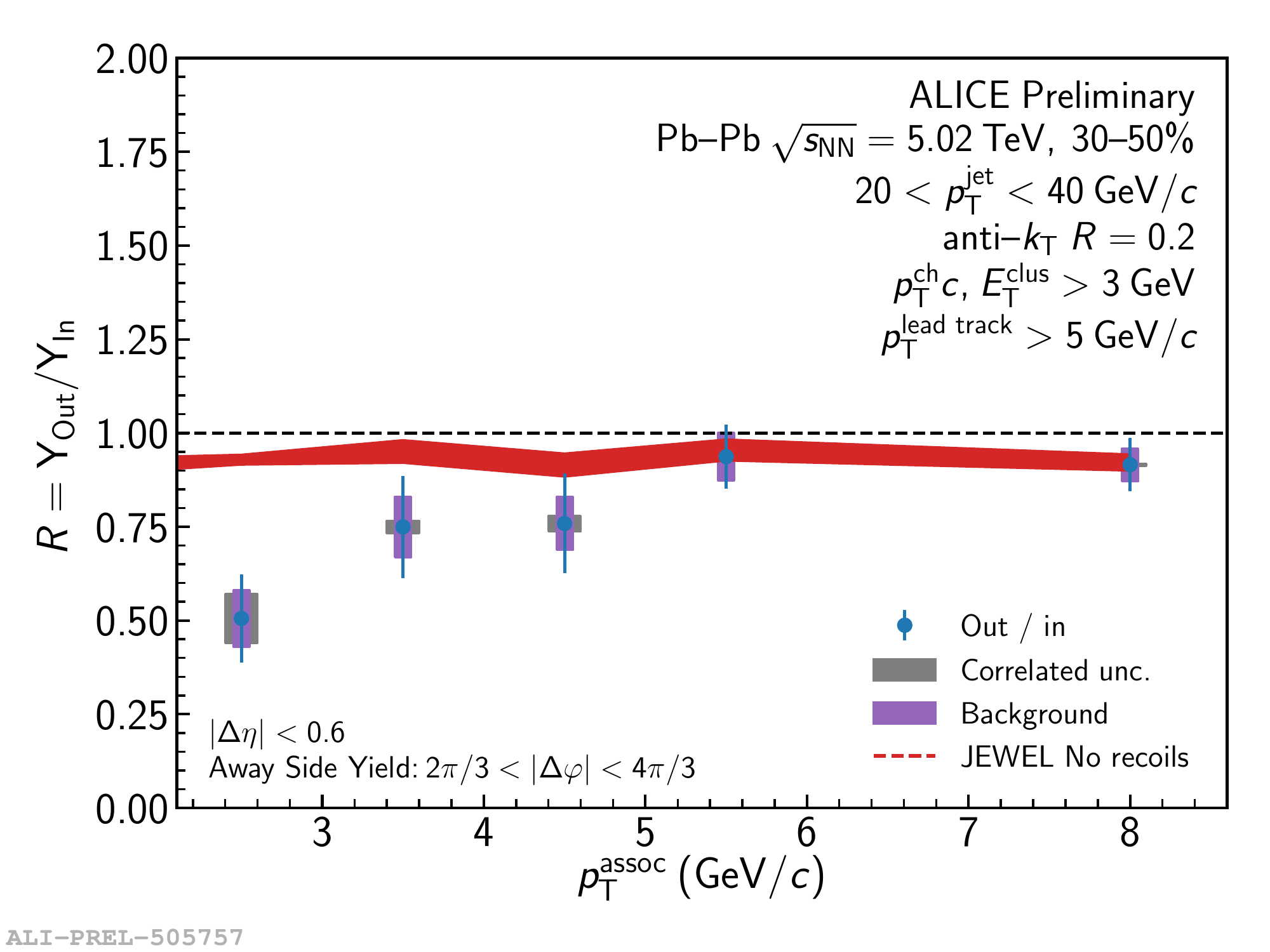}}
\caption{Near-side (left) and away-side (right) jet-hadron correlation yield ratios for out-of-plane to in-plane associated hadrons in Pb--Pb collisions at $\sqrt{s_{\rm NN}} = 5.02$ TeV compared with JEWEL calculations.}
\label{Fig:F2H}
\end{figure}

\vspace{-5mm}

\section{Outlook}
In these proceedings, multiple jet-type observables have been shown with varying sensitivities to the path-length of the traversed medium. While these results may naively appear in conflict, they highlight the importance of understanding what physics an observable accesses before trying to interpret a result. Here, it can be seen that the observables that are most sensitive to path-length differences (in this case, the event-shape engineered jet spectra and jet-particle $v_{\rm 2}$) show results that are consistent with a picture of path-length dependent energy loss. Even in these cases, however, more work remains to be done to better understand these measurements. These results will benefit from the data sample expected to be collected in the LHC Run 3 and the recent upgrade of the ALICE detector. Thus the event-shape engineered jet spectra and high-$p_{\rm T}$ correlations will be measured more differentially and to higher precision in order to better constrain these phenomena.


\vspace{-5mm}

\bibliography{refs}

\end{document}